\documentclass[twocolumn]{aastex61}
\pdfoutput=1 
\usepackage{amsmath,amstext}
\usepackage[T1]{fontenc}
\usepackage{apjfonts} 
\usepackage[figure,figure*]{hypcap}

\shorttitle{Post Transitional Disk Stability}
\shortauthors{Bowens et al.}

\begin{document}
\title{Longterm Stability of Planetary Systems formed from a Transitional Disk}

\author[0000-0003-0949-7212]{Rory Bowens}
\altaffiliation{rpbowens@umich.edu}
\affiliation{Department of Astronomy \& Astrophysics, The Pennsylvania State University, State College, PA, USA}
\affiliation{Center for Exoplanets and Habitable Worlds, The Pennsylvania State University, State College, PA, USA}
\affiliation{Department of Astronomy, The University of Michigan, Ann Arbor, MI, USA}

\author[0000-0002-0711-4516]{Andrew Shannon}
\affiliation{LESIA, Observatoire de Paris, Universit\'{e} PSL, CNRS, Sorbonne Universit\'{e}, Universit\'{e} de Paris, 5 place Jules Janssen, 92195 Meudon, France}
\affiliation{Department of Astronomy \& Astrophysics, The Pennsylvania State University, State College, PA, USA}
\affiliation{Center for Exoplanets and Habitable Worlds, The Pennsylvania State University, State College, PA, USA}

\author[0000-0001-9677-1296]{Rebekah Dawson}
\affiliation{Department of Astronomy \& Astrophysics, The Pennsylvania State University, State College, PA, USA}
\affiliation{Center for Exoplanets and Habitable Worlds, The Pennsylvania State University, State College, PA, USA}

\author[0000-0002-3610-6953]{Jiayin Dong}
\altaffiliation{Flatiron Research Fellow}
\affiliation{Department of Astronomy \& Astrophysics, The Pennsylvania State University, State College, PA, USA}
\affiliation{ Center for Exoplanets and Habitable Worlds, The Pennsylvania State University, State College, PA, USA}
\affiliation{Center for Computational Astrophysics, Flatiron Institute, 162 Fifth Avenue, New York, NY 10010, USA}

\begin{abstract}
  Transitional disks are protoplanetary disks with large and deep central holes in the gas, possibly carved by young planets. 
  Dong, R., \& Dawson, R. 2016, ApJ, 825, 7 simulated systems with multiple giant planets that were capable of carving and maintaining such gaps during the disk stage. Here we continue their simulations by evolving the systems for 10 Gyr after disk dissipation and compare the resulting system architecture to observed giant planet properties, such as their orbital eccentricities and resonances. We find that the simulated systems contain a disproportionately large number of circular orbits compared to observed giant exoplanets. Large eccentricities are generated in simulated systems that go unstable, but too few of our systems go unstable, likely due to our demand that they remain stable during the gas disk stage to maintain cavities. We also explore whether transitional disk inspired initial conditions can account for the observed younger ages of 2:1 resonant systems orbiting mature host stars. Many simulated planet pairs lock into a 2:1 resonance during the gas disk stage, but those that are disrupted tend to be disrupted early, within the first 10 Myr. Our results suggest that systems of giant planets capable of carving and maintaining transitional disks are not the direct predecessors of observed giant planets, either because the transitional disk cavities have a different origin or another process is involved, such as convergent migration that pack planets close together at the end of the transitional disk stage.
\end{abstract}

\section{Introduction}
Planets form in disks of gas and dust that surround young stars, known as protoplanetary disks.  Protoplanetary disks are sometimes observed with a deep and wide gap in the dust distribution (e.g., \citealt{1989AJ.....97.1451S}). These disks are called transitional disks \citep{pietu}.  About $10\%$~of protoplanetary disks are transitional disks \citep{2010ApJS..186..111L}, but it remains unclear whether this fraction indicates that $\sim 10\%$~of protoplanetary disks spend most of their lives as transitional disks or that most protoplanetary disks spend $\sim 10\%$ of their lives as transitional disks \citep{2016PASA...33....5O}.  Numerous theories have been proposed to explain the origins of the gaps \citep[e.g.,][]{2005A&A...434..971D,2007NatPh...3..604C,2007A&A...462..977K,2009AIPC.1158..161S,2015A&A...573A...5V}; the leading theories are photoevaporation \citep{2001MNRAS.328..485C,2006MNRAS.369..216A,2006MNRAS.369..229A,2011MNRAS.412...13O,2012MNRAS.422.1880O} and planetary sculpting \citep{2005ApJ...630L.185C,dods11,zhu11,2012ApJ...755....6Z,2015ApJ...809...93D}. Photoevaporative winds can deplete the inner disk when the photoevaporative mass loss rate exceeds the accretion rate. Although early photoevaporation models (e.g., \citealt{2011MNRAS.412...13O,2012MNRAS.422.1880O}) produced photoevaporative mass loss rates too low to be consistent with the high observed accretion rates in transitional disks \citep{ercl17}, recent models (e.g., \citealt{ercl21,pigl21}) that more accurately model the temperature structure of the disk can meet or exceed half of observed accretion rates. Transitional disks with higher accretion rates may be the consequence of stronger photoevaporative winds in carbon-depleted disks \citep{ercl18,wolf19}. Alternatively, they may be the result of magnetically driven supersonic accretion flows (\citealt{wang17}; but see \citealt{ercl18} for observational arguments against this hypothesis). More in-depth reviews of the observational properties of transitional disks and theory of their origins can be found in \citet{2014prpl.conf..497E,2016PASA...33....5O,2017RSOS....470114E,2017ASSL..445...39V}.

In this work, we focus on the question of whether gaps in transitional disks form by planetary sculpting \citep{1984ApJ...285..818P,1993ApJ...405L..71M,2004A&A...425L...9P}. Planetary sculpting is a process in which planets or brown dwarfs formed in the protoplanetary disk clear a gap (\citealt{dods11,zhu11}, \citealt{rose20}; see \citealt{paar22} for a recent review). This theory is supported by the discovery of protoplanets forming in the gap of the transitional disk PDS 70 \citep{2018A&A...617A..44K,2019NatAs.tmp..329H,2019arXiv190606308I,2019mule}, possibly LkCa 15 (\citealt{2012ApJ...745....5K,2015Natur.527..342S}; but see also \citealt{2019ApJ...877L...3C}), 
and a few other candidates \citep[HD 100546, HD 142527, HD 169142;][]{2013ApJ...766L...1Q,2012ApJ...753L..38B,2014ApJ...792L..23R}. Further supporting this hypothesis, sub-structures in proto-planetary disks have also been attributed to planetary sculpting (e.g., \citealt{huan18,long18,zhan18,10.1093/mnras/stab3503}). Although the sculpting planets creating the deep and wide gaps in transitional disks are typically assumed to be Jovian -- and we will make that assumption here -- \citet{fung2017} showed that compact configurations of super-Earths can clear inner cavities in disks with very low viscosity. See \citet{ginz18} and \citet{garr22} for more on the properties of such gaps.

To account for the gaps seen in transitional disks, planetary systems must remain stable over the observed disk timescale \citep{Tamayo_2015}.  To assess which planetary system architectures are capable of producing transitional disks, \citet{dong16} (hereafter DD16) performed $N$-body simulations of such planetary systems, using three to six giant planets spaced from 3 to 30 AU. The depletion of the gaps the planets  produced was based on different assumptions for the disk viscosity and scale height, and these depletions dictated the spacings of the planets necessary to produce a continuous gap. DD16 included an eccentricity damping force from the gas in the gap. They found that a subset of planetary system configurations remained stable over a typical one million year disk lifetime. Thus they concluded it was plausible that the subset of planetary systems that contain Jovians planets in the 3 to 30 AU range all appear as transitional disks during the protoplanetary stage. Under this hypothesis, the phenomenon of transitional disks is not pervasive, affecting all protoplanetary disks for $\sim$10\% of their lifetime, but instead is restricted to the $\sim$10\% of disks that happen to harbor giant planets.

Because DD16 conducted simulations only during the gas disk stage, we cannot directly compare those systems to mature planetary systems observed around main sequence field stars. After the protoplanetary disk dissipates in $\lesssim 10^7$~years \citep{2001ApJ...553L.153H,2014ApJ...793L..34P}, the systems may go unstable, as systems of massive planets in such compact configurations often do \citep{1996Icar..119..261C,2009Icar..201..381S,2016ApJ...823..118M}. However, the damping during the protoplanetary disk state may allow these systems to find long-term stable compact dynamical states (e.g., \citealt{1996MNRAS.280..854M,2002ApJ...567..596L,daws16,2020ApJ...904..157M}).  The four tightly packed jovian planets around HR 8799 \citep{2008Sci...322.1348M,2010Natur.468.1080M} may be in such a dynamical state \citep{2010ApJ...710.1408F,2014MNRAS.440.3140G,2018AJ....156..192W}, and thus perhaps an example of the post-gas evolution of such systems. 

Here, we seek to understand the longer-term dynamical evolution of the giant planet planetary systems capable of sculpting deep and wide gaps in  transitional disks. \citet{stx772} performed a case study of HL Tau along these lines, finding ejection of one or more planets to be the most common outcome.  How this result can be generalized remains an open question. Previous works simulating the longterm evolution of planetary systems compared simulated systems' eccentricity and semimajor axis distributions \citep[e.g.,][]{2008ApJ...686..580C,2008ApJ...686..603J,2009MNRAS.394L..26M,2014ApJ...786..101P,2019A&A...629L...7C} to those of observed planets \citep[e.g.,][]{2011arXiv1109.2497M,2012ApJ...756..122D,2015ARA&A..53..409W,2016PNAS..11311431X}. However, those works typically began with ad hoc tightly packed initial configurations of orbits to quickly induce instabilities.  Here, we use initial conditions physically motivated from DD16, as our interest is in how well those hypothesized transitional-disk sculpting systems match observed systems. We also explore the prevalence and evolution of mean motion resonances in these simulated systems, which were common during the stage simulated by DD16. Motivated by \citet{koriski}'s finding that 2:1 mean motion resonance (MMR) systems are younger on average, we will assess whether the systems that begin in MMR will break within observable timescales.

We describe our simulations in Section \ref{section:methods}.  We identify the presence and behavior of orbital resonances Section \ref{sec:res}. We assess longterm stability and which characteristics affect  it in Section \ref{sec:stability}. We investigate the planets' eccentricities and compare to  those of observed exoplanets in Section \ref{sec:eccentricity}. We present our conclusions in Section \ref{sec:conclusion}.

\section{Simulations}
\label{section:methods}
We simulate the long-term evolution of the multi-planet systems that are capable of opening the gaps observed in transitional disks. Most of our simulations start where DD16's left off, at the end of transitional disk stage, with gas damping forces turned off, which is an approximation that the gas disk is immediately removed. Their final planet masses, positions, and velocities from the gas stage are our initial post-gas conditions. However, since we do run some additional gas stage simulations, we describe the gas stage simulations in detail in Appendix A.

For our post-gas simulations use the {\tt mercury6} Bulirsch-Stoer integrator with a 1000 AU ejection distance, a solar mass and solar radius central body, an accuracy parameter of $10^{-12}$, and medium precision outputs every million years. Our approximation of the instantaneous removal of the gas disk does not induce problems for several reasons: 1) as we will show the majority of systems remain stable after the removal, 2) the gas surface density was low to begin with due to depletion in the gas, and 3) simulations show that instantaneously removing a depleted gas disk does not significantly alter the resonant dynamics \citep{2020ApJ...904..157M}. 

The configurations are summarized in Table \ref{tab:config}. Configuration names are defined as planet number followed by planet mass in Jupiters (with letters used for configurations with the same planet numbers and masses). Being continuations of DD16, the systems had between 3 and 6 equal mass planets, where the number of planets required was dictated by the gap widths. We did not include configurations that DD16 found to be unsuitable for creating proto-planetary disk cavities because the amount of gas expected to be present in the cavity would drive the planets too far apart via resonant repulsion to create overlapping gaps (their 4-10acd, 5-5ac) or would be insufficient to stabilize the configuration during the gas disk stage (their 6-2). For each configuration, DD16 used various assumed gas surface densities since ALMA gas observations and chemical disk modeling are unable to constrain gas depletion in the inner few AU of a disk. For each configuration, we use the simulations with the gas surface density consistent with the expected depletion in the cavity  (reported in \citealt{dong16} based on \citealt{fung14}). For their gas stage simulations, DD16 ran 10 random realizations for each configuration. We supplement with an additional 10 gas stage realizations for each configuration, that we then continue in the post-gas stage. See Appendix A for more details about the gas stage simulations. In summary, we ran eighteen realizations of twenty variations of the disk initial conditions for 10 Gyr, for a total of 360 simulations. We found that 14 simulations had gone unstable during the gas disk stage (i.e., lost a planet via ejections or collisions); those simulations are considered to have instability events at 0 Gyr in all future analysis. We also found that two simulations ended the 10 Gyr simulations with zero planets due to central body collisions.

\begin{table}
\centering
\caption{Summary of 3 -- 30 AU Systems}
\footnotesize
\begin{tabular}{l|l|l|l|l|lll|l|l}
\\
\hline
Name & $\Sigma_{30}$ & $\Delta_0$ & GDI & 2:1 MMR \% & 10 Gyr \\
& g cm$^{-2}$&($R_H$)& Sims. & (Near \%)  & Stability \% \\
\hline
{\tt 3-5}&	0.001& 6.2 & 0 & 15, (20) & 100\\ 
{\tt 3-10a}	&0.0001 & 5.2 & 0 & 40, (40) & 100	\\ 
 {\tt 3-10b}&0.1  & 4.8 & 0 & 75, (25) & 100\\ 
 {\tt 3-10c}&0.0001 & 4.4 & 0 & 40, (25) & 95\\ 
 {\tt 3-10d}&	0.1 & 4.4 & 0 & 75, (25) & 100\\ 
 {\tt 3-10e}	&	0.0001 & 4.0 & 1 & 0, (25) & 80\\ 
 {\tt 4-2}	&	0.001 &	6.1 & 0 & 85,  (15) & 95\\ 
{\tt 4-5a}	&	0.001 & 4.7 & 0 & 85,  (15) & 95\\ 
{\tt 4-5b}	&	0.1 & 4.6 & 0 & 20,  (80) & 100\\ 	
{\tt 4-5c}	&	0.1	 & 4.6 & 0 & 20, (75) & 100\\ 
{\tt 4-5d}	&0.001 & 4.0& 1 & 45, (45) & 20  \\ 
{\tt 4-10b}	&	0.01 & 3.9 & 2 & 40,  (35) & 40\\ 
 {\tt 5-1}	&	0.1	 & 6.2 & 0 & 60,  (40) & 90\\ 
{\tt 5-2a}	&	0.1 & 4.6 & 0 & 95,  (5) & 30\\ 
{\tt 5-2b}	&	0.1	& 4.2 & 7 & 65, (10) & 30\\ 
 {\tt 5-5b}	& 0.01 & 3.6 & 1 & 70, (25) & 40\\ 
 {\tt 6-0.5}	&	1&	5.6 & 0 & 50,  (40) & 5\\ 
 {\tt 6-1}	&	1	  & 4.4 & 2 & 45,  (25) & 5\\ 
\hline
\end{tabular}
\label{tab:config}
\tablecomments{The parameter $\Sigma_{30}$ indicates normalization the gas surface density inside the depleted gap for the gas stage simulation (Appendix A), where $\Sigma_{\rm gas}=\Sigma_{30} \left(\frac{a}{30\rm AU}\right)^{-3/2}$, and a normalization of $\Sigma_{30} = $10 g cm$^{-2}$ corresponds to the minimum mass solar nebula. GDI Sims. gives the number of simulations (out of the twenty in a set) that experienced instabilities during the gas phase integration. These simulations were still run for the following 10 Gyr integration with their surviving planets.}
\end{table}

To better compare giant planets discovered by the radial velocity method -- which are commonly observed at $\sim$ 1--3 AU -- we ran an additional set of simulations that added an interior planet. We refer to the original set as 3--30 AU simulations and additional set as 1--30 AU simulations. The additional planet is placed interior to the others by the average Hill spacing of original configuration. First the average Hill spacing of the original configuration is determined by averaging every pair's separation in all \textasciitilde20 simulations for the configuration, excluding those that went unstable during the gas disk stage. Then the position for the innermost planet is determined relative to the initial location of the 3 AU planet according to the average Hill spacing. The newly created 1 AU planet is appended to the corresponding original 3--30 AU simulation during the gas stage, in order to make the final results more comparable. Similar to the 3--30 AU gas stage simulations (Appendix A), these systems are simulated for 1 Myr with gas damping present using the hybrid integrator (timestep of 3 days, accuracy parameter of $10^{-12}$). However, they are then simulated for only 1 Gyr with gas damping shut off using the Bulirsch-Stoer integrator. This reduced simulation time is necessary for the efficient completion of the simulations.

\section{Resonant behavior}
\label{sec:res}
Here we identify systems containing planets in two-body and three-body resonances. We assess resonance only for adjacent planet pairs (or triplets in the case of three body resonances) and by looking for librating resonant angles.  For example, for the 2:1 resonance, the resonant angles are $\phi_{\rm{in}}$~and $\phi_{\rm{out}}$:
\begin{equation}
\phi_{in} = 2 \lambda_{in} - \lambda_{out} - \varpi_{in}
\end{equation}
and 
\begin{equation}
\phi_{out} = 2 \lambda_{in} - \lambda_{out} - \varpi_{out}
\end{equation}
where $\lambda_{\rm{in}}$~and $\lambda_{\rm{out}}$~are the mean longitude of the inner and outer planet, respectively, and $\varpi_{\rm{in}}$~and $\varpi_{\rm{out}}$~are the longitude of pericenter of the inner and outer planets, respectively. To ensure the resonant angles are well sampled, we run simulations with more frequent output (every 10 years) for $10^5$ years, using the same starting point as the standard simulations (i.e., immediately after gas disk stage). The 3--30 AU systems are run with a Bulirsch-Stoer integrator ($10^{-12}$ accuracy parameter) while the 1--30 AU are run with a hybrid integrator (3 day timestep, $10^{-12}$ accuracy parameter).  

\begin{figure}[ht]
\centering
\includegraphics[width=.4\textwidth]{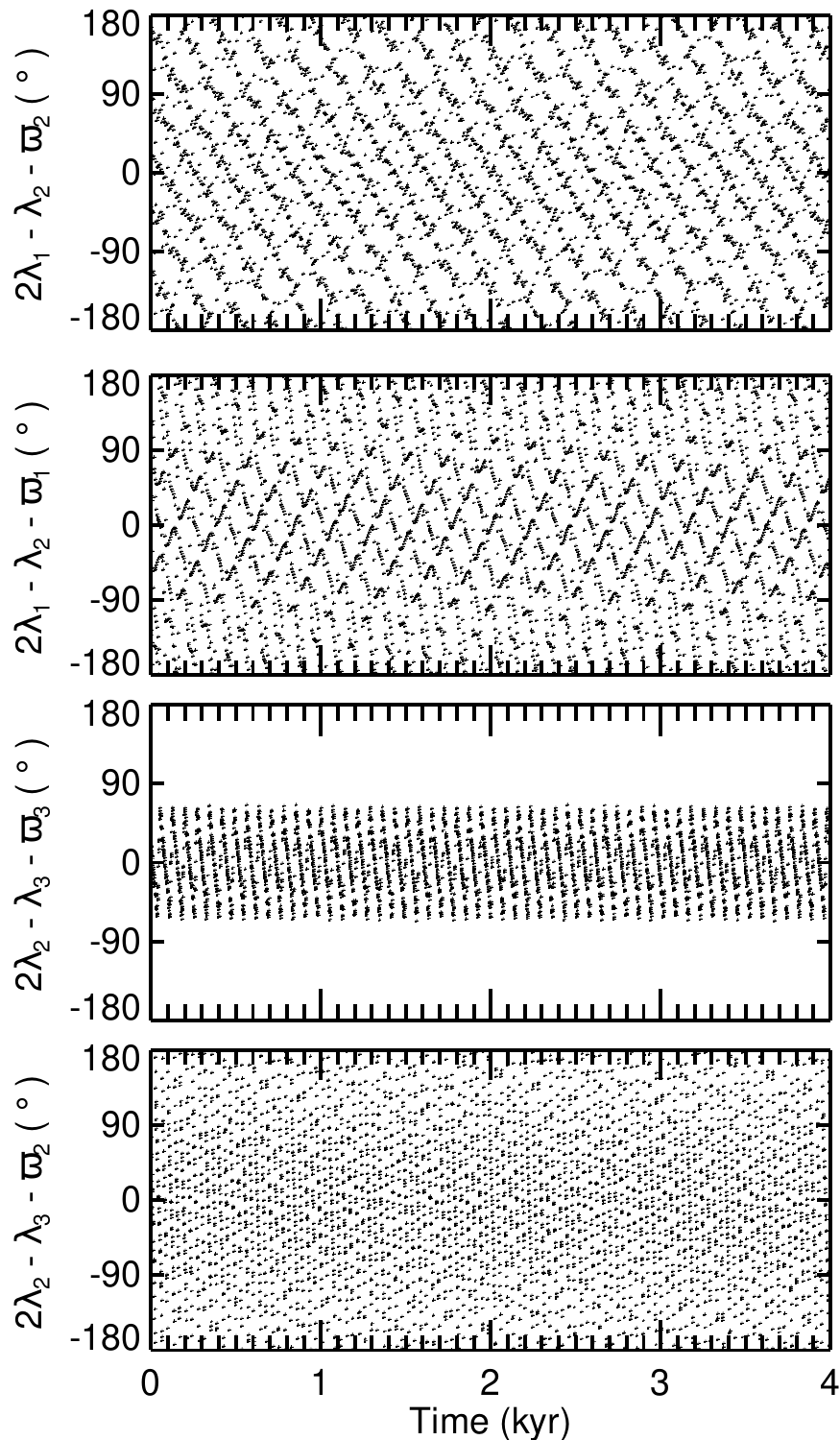}
\caption{\label{fig:true_res_example}
The resonance angle for planet pairs in a simulation from Config {\tt 3-10b}, starting in the post-gas stage. Planets 2 and 3 (the inner two of the three; panel 3) happened to begin the gas disk stage in resonance and maintained this configuration in the post-gas stage. In other cases, our simulated planets get captured into resonance during the gas disk stage.}
\end{figure}

The libration centers for the massive planets in our simulations are 0 or 180$^\circ$ \citep{dong16}. We consider a planet pair resonant with libration about 0 if the resonant angle does not come within $\pm$0.5$^\circ$ of 180$^\circ$ during the simulation (Fig. \ref{fig:true_res_example}). Likewise, we consider a planet pair resonant with libration about 180$^\circ$ if the resonant angle does not come within $\pm$0.5$^\circ$ of 0 during the simulation.
 
\begin{figure}[ht]
\centering
\includegraphics[width=.4\textwidth]{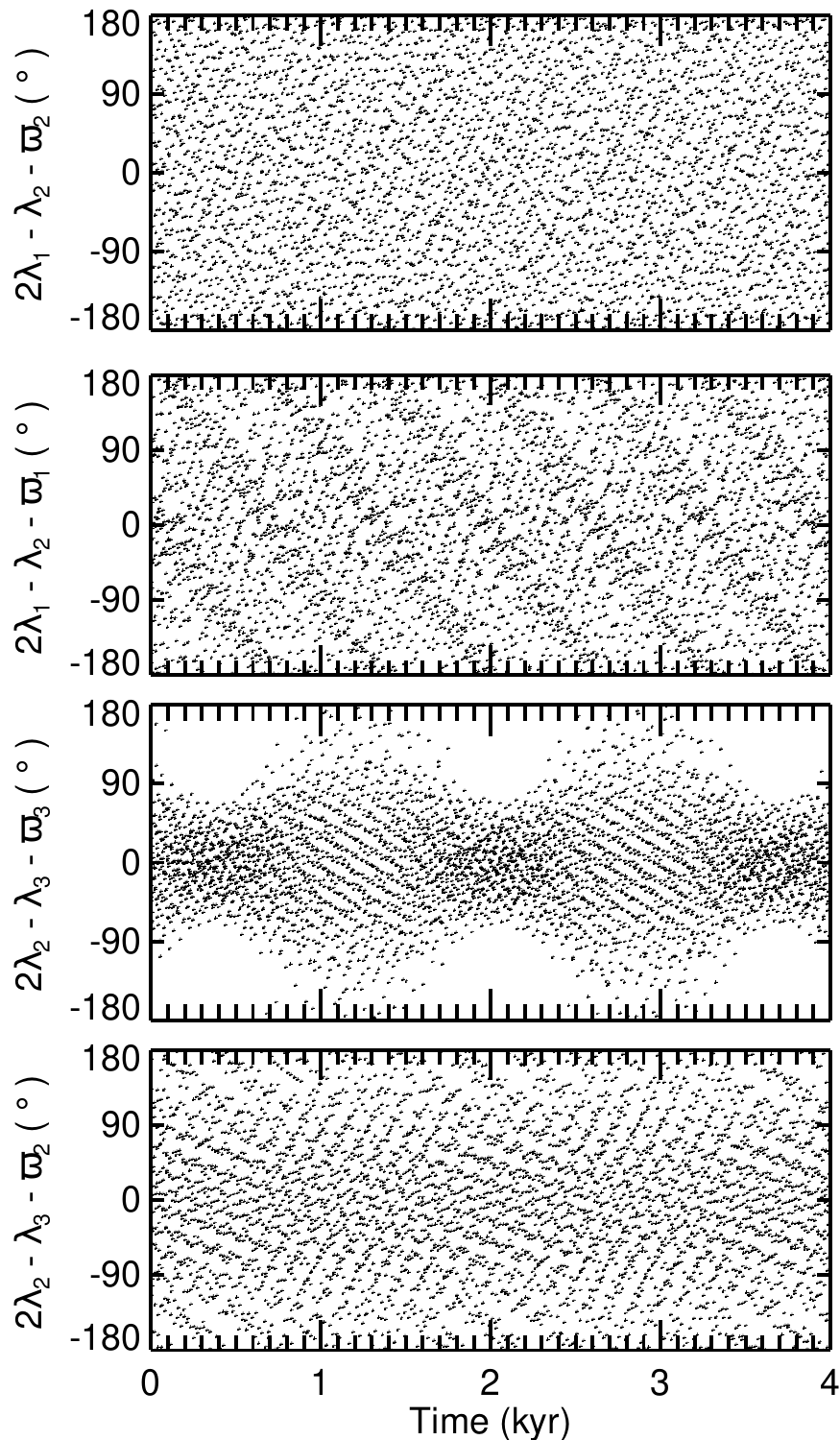}
\caption{\label{fig:near_res_example}
The resonance angle among planet pairs in a simulation from Config {\tt 3-10a}, starting in the post-gas stage. Each resonance angle circulates but we consider planets 2 and 3 (the inner two of the three; panel 3) to be ``near resonance'' because the angle lingers within $\pm90^\circ$ of 0.The pair reached this configuration part way through the earlier gas stage.
}
\end{figure}

Furthermore, we label systems with at least one planet pair whose resonant angle remained within $\pm$170$^\circ$ of 0 or 180 for 97.5\% of the time as near resonance. These angles linger near a specific value as they circulate (Fig. \ref{fig:near_res_example}). If the resonant angles circulate without lingering for all the planet pairs in a system, we classify the system as non-resonant. We inspect a subset of 54 of the resonant angle plots by eye to ensure the classification was reliable and that 0.5$^\circ$ worked well as a cut-off with our sampling frequency. We also considered a stricter resonant angle range, requiring it to stay $\pm$50$^\circ$ of 0$^\circ$ or 180 $^\circ$ during the simulation. This did reduce the number of resonant systems by 12\% but had no significant impacts on other results.

Our resonance identification approach can fail for systems that go unstable quickly. To solve this problem, we truncate the resonance data at 75\% of the instability timescale, defined as the timescale when a collision or ejection occurs (e.g., for a system that went unstable after four hundred thousand years, we use the resonant angle during the first three hundred thousand years). We label all planets that collided or were ejected before 60,000 years as having undergone "rapid instability." Less then ten systems had rapid instability so this classification is only a minor part of the results. Some systems that went unstable during the gas disk stage only had a single planet remaining at the start of the gas-free stage. These systems are likewise given a unique label apart from resonance or no resonance.

We find that it is more likely for the inner planets to be in 2:1 MMR than the outer planets. Approximately 50\% of the innermost planet pairs are in resonance in all planet systems. For outermost pairs, 14\% of those in the five planet systems are in 2:1 MMR, 3\% of the outer pairs in the six planet systems, and none of those in the three or four planet systems. The higher fraction of true resonance among inner pairs vs. outer may be because the higher surface gas surface density and shorter orbital timescales in the inner disk facilitate resonance capture during the gas disk stage.

In our four, five, and six planet systems, outer pairs are near 2:1 MMR (``lingering'' systems with at least one planet pair whose resonant angle remained within $\pm$170$^\circ$ of 0 or 180 for 97.5\% of the time), in approximately 70\% pairs while inner pairs experience closer to 30\% with near 2:1 MMR. Intermediate planet pairs show intermediate rates of near 2:1 MMR. However, in the three planet systems, 30\% of the inner pairs are near resonance but none of the outer pairs are near resonance. 

\begin{figure}[ht]
\centering
\includegraphics[width=.4\textwidth]{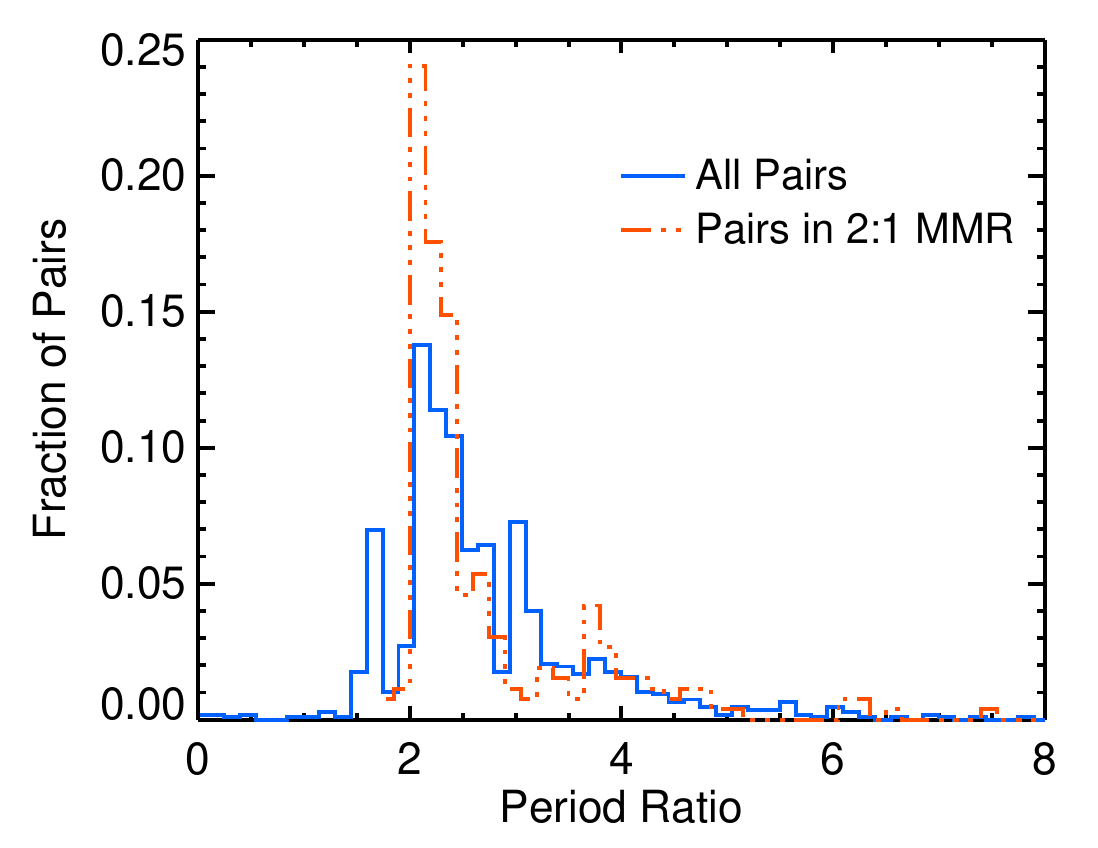}
\caption{\label{fig:ratio_histo_comp}
The ratio of the periods of adjacent planets for all 3--30 AU simulated systems (blue line) and systems in 2:1 MMR (red-dotted/dashed), as defined by libration of the resonant angle. Systems in 2:1 MMR do not all have period ratios near 2. Many have ratios much larger than 2, even up to 6. Similar trends were observed for the 1--30 AU simulations.
}
\end{figure}

We also examine the period ratios of 2:1 MMR systems (Fig.  \ref{fig:ratio_histo_comp}). Consistent with DD16's findings, planets with period ratios far outside of their nominal resonance can have librating resonant angles. There are some systems that are librating in the 2:1 that have a period ratio greater than 2.5, even up to 6 (i.e., their 2:1 resonant angle is librating). Libration of the 2:1 resonant angle at such large period ratios is possible when the longitude of periapse precesses quickly and is caused in our simulations by the eccentricity damping during the gas disk stage. This phenomenon, which is more generally established by a dissipative change to the eccentricity (which may or may not be accompanied by a change in semi-major axis) is known as resonant repulsion (e.g., \citealt{lith12}). We find 47\% of systems that lie within 10\% of a period ratio of 2 are in 2:1 MMR. 

In later sections, we will focus on the 2:1 MMR because other two body resonant angles rarely librate in our simulations (e.g., 3:2, 3:1, 4:3). Among the 360 simulations, we find four systems contain one or more pairs that librate in the 3:2 resonance (twenty-five near resonance), one system in the 3:1 resonance (one-hundred sixty-nine near-resonance), and zero systems in the 4:3 resonance. We also examined various three body resonances. First, we define the three body librating angle equation:
\begin{equation}
\phi_{3b/p,q}(1,2,3) = p\lambda_1 - (p + q)\lambda_2 + q\lambda_3
\end{equation}
In the above, $1$, $2$, and $3$ refer to the outer, middle, and inner planet, respectively. We examined multiple potential resonances and found that only 9:6:4, 15:12:8, 4:2:1, and particularly 3:2:1 had a significant number of three body resonance cases. 22\% of systems displayed at least one type of three body resonance (1\% for three planet systems, 23\% for four planet systems, 45\% for five planet systems, and 35\% for six planet systems). In all cases, there were less than 5 near resonance simulations. 

\section{Stability}
\label{sec:stability}

In approximately 31\% of the simulations, an ejection or collision occurs over 10 Gyr. We define a system as stable if none of the member planets were ejected or collided during either the gas or post-gas stage. 

\subsection{Impact of Planet Number and 2:1 MMR}
Our data consists of 360 3--30 AU systems. Each set of 20 systems is defined by a planet number and initial gas surface density profile (Table \ref{tab:config}). In total there are 6 three planet sets, 6 four planet sets, 4 five planet sets, and 2 six planet sets.

We find that systems with more planets experience more collisions and/or ejections. Examining systems that remained stable during the gas disk stage, approximately 13\% of the outermost planets are lost (ejected or collided) during the 10 Gyr gas-free phase in contrast to 20\% of inner planets lost. The difference is mostly due to more collisions among inner planets. In six planet systems, approximately 33\% of planets were ejected and 15\% collided. These values are significantly higher than the averages for the four planet systems (9\% ejected and 3\% collided). We conclude that additional planets reduce a system's stability but that origin of the ejected planets is fairly uniform across the semimajor axis range.

\begin{figure}[ht]
\centering
\includegraphics[width=.4\textwidth]{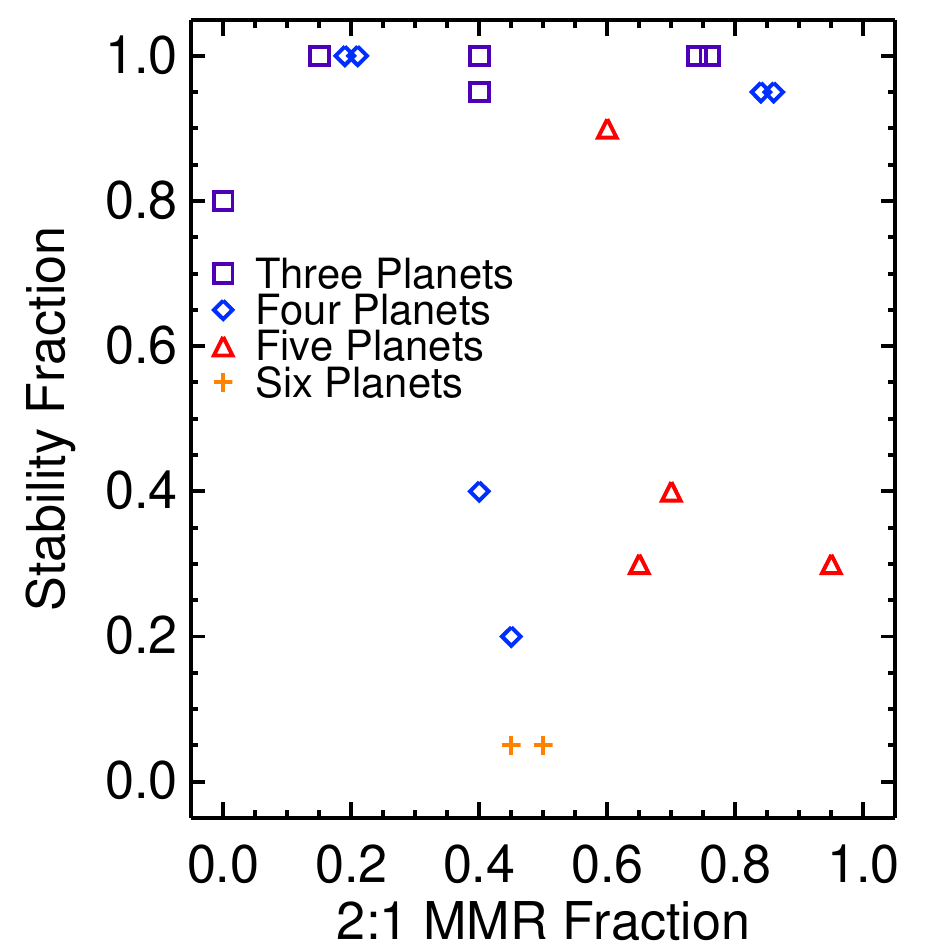}
\caption{\label{fig:mmr_stability}
The fraction of the 3--30 AU simulations with at least one 2:1 MMR in a configuration versus the overall stability of the configuration. Each configuration contains 20 simulations. For readability, the three pairs of identical sets (3-10b and 3-10d, 4-2 and 4-5a, 4-5b and 4-5c) have had their 2:1 MMR \% (75\%, 85\%, and 20\%, respectively) split slightly.
}
\end{figure}

\label{2_1_mmr_stability}
We plot the initial fraction of 3--30 AU systems in 2:1 MMR against the stability fraction, color coded by planet count (Fig. \ref{fig:mmr_stability}). Recall that for a given planet number, different sets have different planet masses and spacings. The initial fraction of systems in resonance alone does not appear to impact the stability but the initial planet count does. Moreover, the fraction of systems in resonance is not strongly correlated with the initial planet count, so it appears that planet count is independently driving the stability. The 3 planet systems remain stable \textasciitilde90\% of the time. The four and five planet systems display a range of stability from low (\textasciitilde20\% of a set stable) to high (\textasciitilde95\% of a set stable) values. Both sets of six planet systems have a low stability fraction (5\%). 

\begin{figure}[ht]
\centering
\includegraphics[width=.45\textwidth]{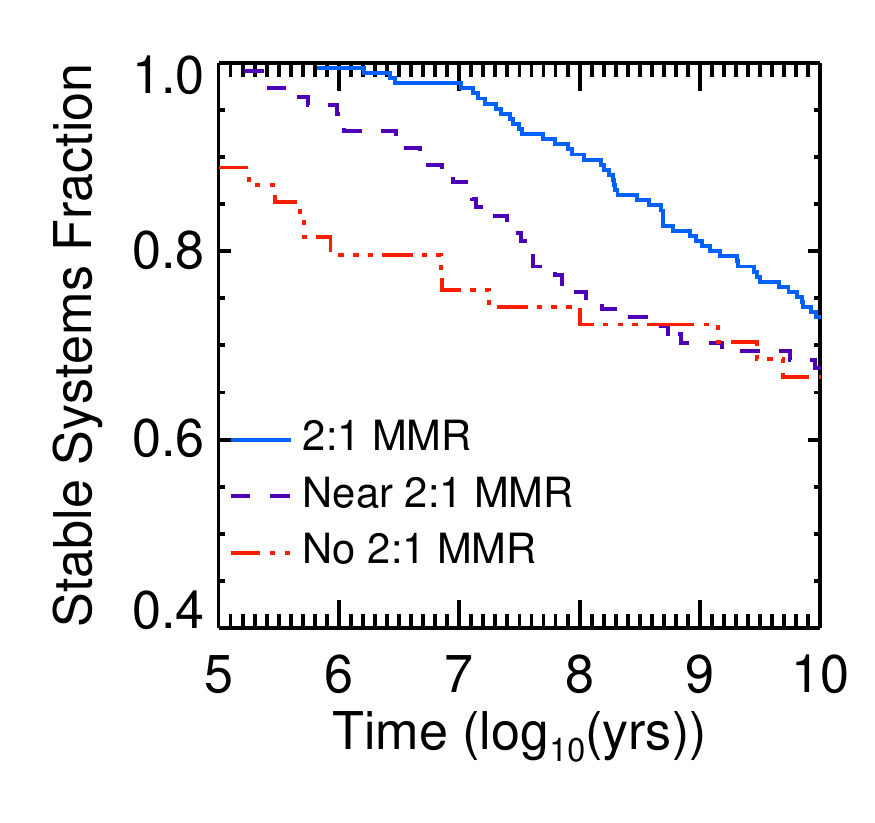}
\caption{\label{fig:stability_lifetime_histogram}
A CDF comparing instability times based on whether the system contains resonant, near-resonant, or no resonant planets. There are 185 resonant systems (of which 135 or 73\% were stable), 111 near resonant systems (of which 75 or 68\% were stable), and 54 no resonant systems (of which 36 or 67\% were stable). Non-resonant systems start below 100\% because some simulations went unstable during the gas disk stage and are marked as unstable at 0 Gyr. Notably, there is little discrepancy between the final stability rates, suggesting the presence of 2:1 MMR does not significantly impact overall stability.
}
\end{figure}

We define the stability timescale as the time until any ejection or collision occurs within the system. We find some systematic differences in stability timescale for systems with 2:1 MMR and without (see Figure \ref{fig:stability_lifetime_histogram}). The timescales at which systems went unstable differs slightly between the three MMR categories (resonance, near resonance, and no resonance). However, the overall rate of instabilities by 10 Gyrs between the three categories was similar: 73\% stable, 68\% stable, and 67\% stable for resonance, near resonance, and no resonance; respectively. 10 systems could not have their resonance status determined due to rapid instability and thus were not used in the above assessments. From this data, we conclude that 2:1 MMR can improve the stability of a system in the short term (Myr timescale) but that this advantage is nullified with time. Based on only small differences in the final stability rates, we conclude the presence of one or more 2:1 MMR pairs within a system did not significantly impact the final number of planets that collided or were ejected of 10 Gyr.

\begin{figure}[ht]
\centering
\includegraphics[width=.45\textwidth]{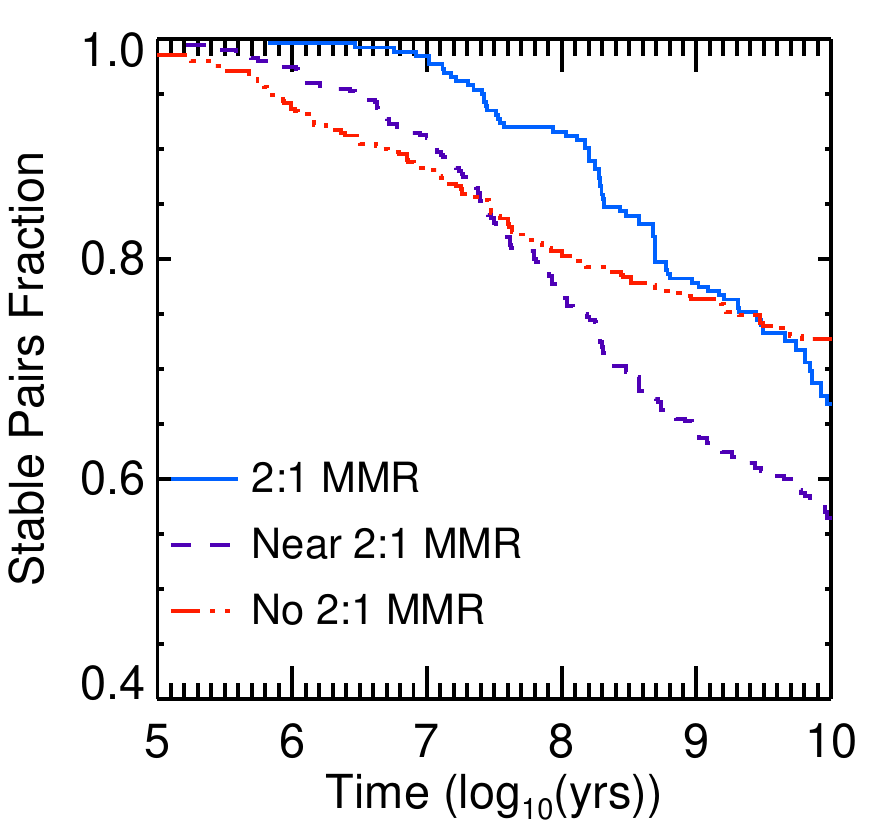}
\caption{\label{fig:stability_lifetime_histogram_pbyp}
A CDF comparing instability times for the three categories of resonance, now plotting per pair rather than per system. There are 262 resonant pairs (of which 67\% were stable), 400 near resonant pairs (of which 57\% were stable), and 411 non- resonant pairs (of which 73\% were stable).
}
\end{figure}

The above analysis of how resonances affect system stability looks at a system level: if at least one pair is in resonance, we consider it a resonant system. However, some trends may only be observable at the individual pair level. In Figure \ref{fig:stability_lifetime_histogram_pbyp}, we report stability timescales in a fashion identical to Figure \ref{fig:stability_lifetime_histogram} but for adjacent pairs rather than systems. A pair is considered stable until one member experiences an instability event. The pairs follow a similar trend to the systems, though we find more significant evidence (compared to the system-level analysis) that near 2:1 MMR pairs are less stable over the long term. The higher instability of near systems may be the result of close spacing without the stabilizing influence of resonance or chaotic behavior at the resonance separatrix.

\begin{figure}[ht]
\centering
\includegraphics[width=.45\textwidth]{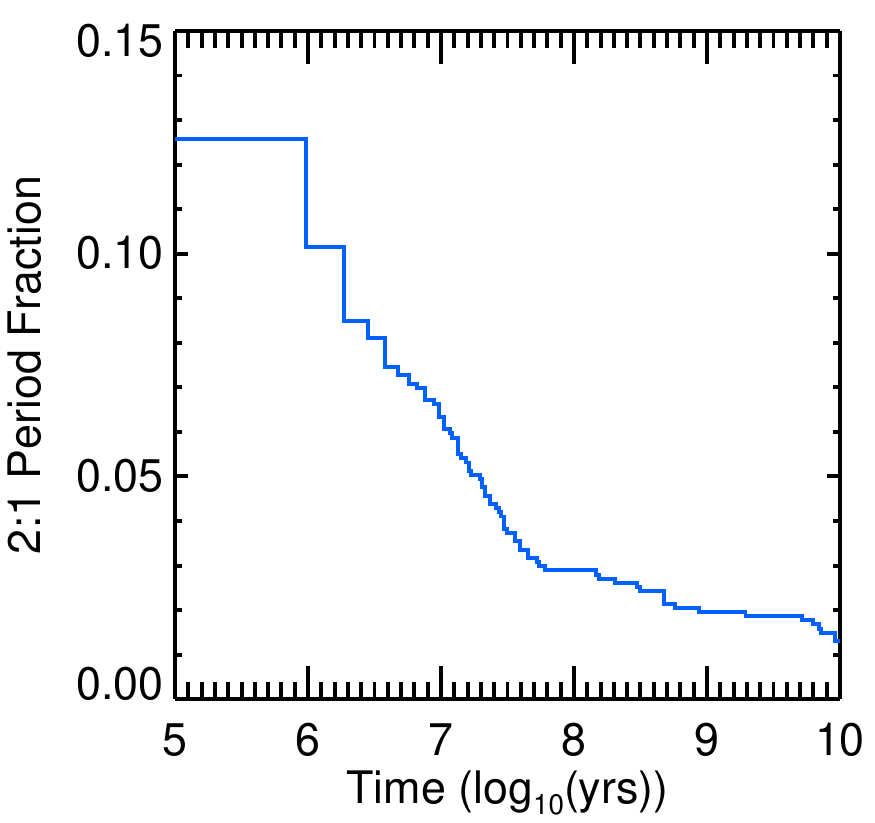}
\caption{\label{fig:stability_lifetime_histogram_period}
A CDF showing the fraction of pairs with 2:1 MMR based on the period ratio. About 13\% of pairs begin in resonance according to this metric.
}
\end{figure}

We also plot a calculation of resonance pairs based on period ratios rather than angle libration (Figure \ref{fig:stability_lifetime_histogram_period}). Identifying systems near integer period ratios allows us to more directly compare to observational  data where the orbital parameters are typically not well-constrained enough to determine if the resonance angle is librating. We determine if a period ratio is resonant according to the criteria in \citet{koriski}:
\begin{equation}
    \delta = 2\frac{|r-r_c|}{r+r_c} \leq 0.1
\end{equation}
where $r$ is the measured period ratio and $r_c$ the resonance period ratio (in this case 2). The output timesteps are every Myr and once a system leaves resonance we do not consider it capable of reentering. About 13\% of our pairs begin in resonance post-gas stage according to this criterion, compared to 24\% using the angle libration criterion. Most of these pairs are in five or six planet configurations, with some sets having up to 40\% of pairs near the 2:1. Although the fraction of pairs near the 2:1 period ratio declines over time, most disruptions occur in  the first 50 Myr. Therefore, this behavior is not a good explanation for the trend of younger ages for resonant pairs found by \citet{koriski}, which requires a typical disruption timescale $\sim$1--10 Gy and near 100\% initial fraction \citep{dong16}.

\subsection{Impact of Three-Body MMR}
\label{three_body_res}
We consider the impact of three-body MMR on the stability of systems. These classifications are once again carried out at the system level. The most common type of three-body MMR in our systems is 3:2:1, occurring in 11\% of systems. We found that \textasciitilde 51\% systems with at least one of the three-body resonances present had a final stability rate, similar to the rate of 55\% of all systems with four or more planets (our three planet systems have a higher stability rate but no three body resonances). The other three body MMRs have small occurrence rates and thus limited statistics, precluding any conclusions on their significance, if any. 

\subsection{Impact of Mutual Hill radii}
\label{hill_stability}

\begin{figure*}[ht]
\centering
\includegraphics[width=.4\textwidth]{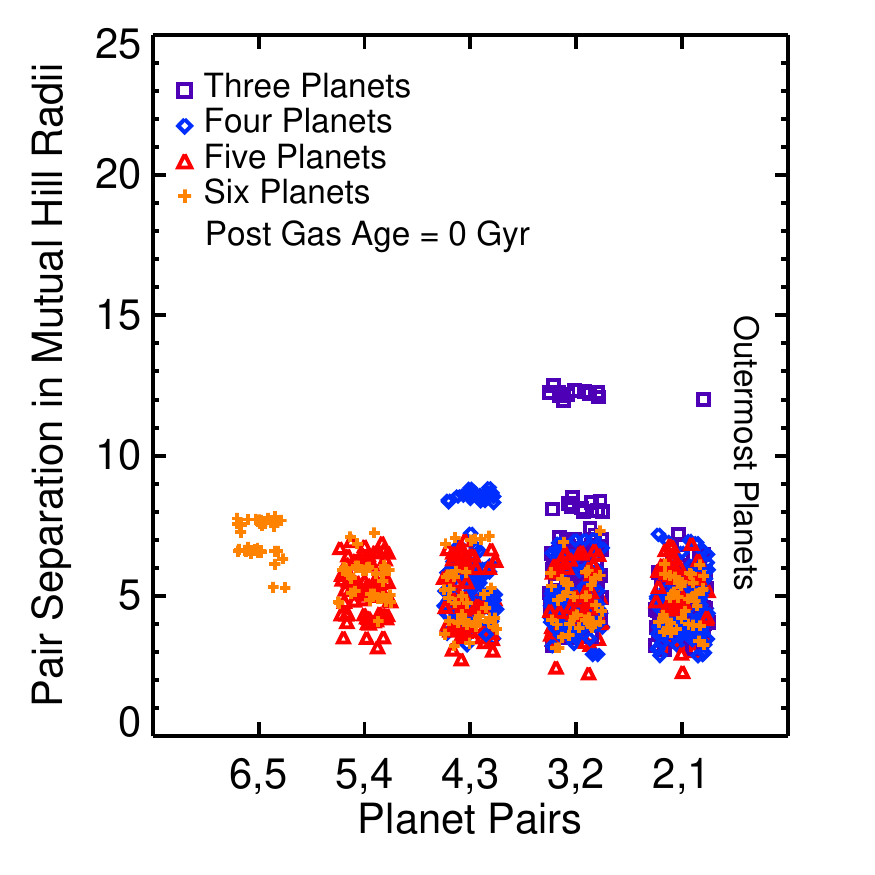}
\includegraphics[width=.4\textwidth]{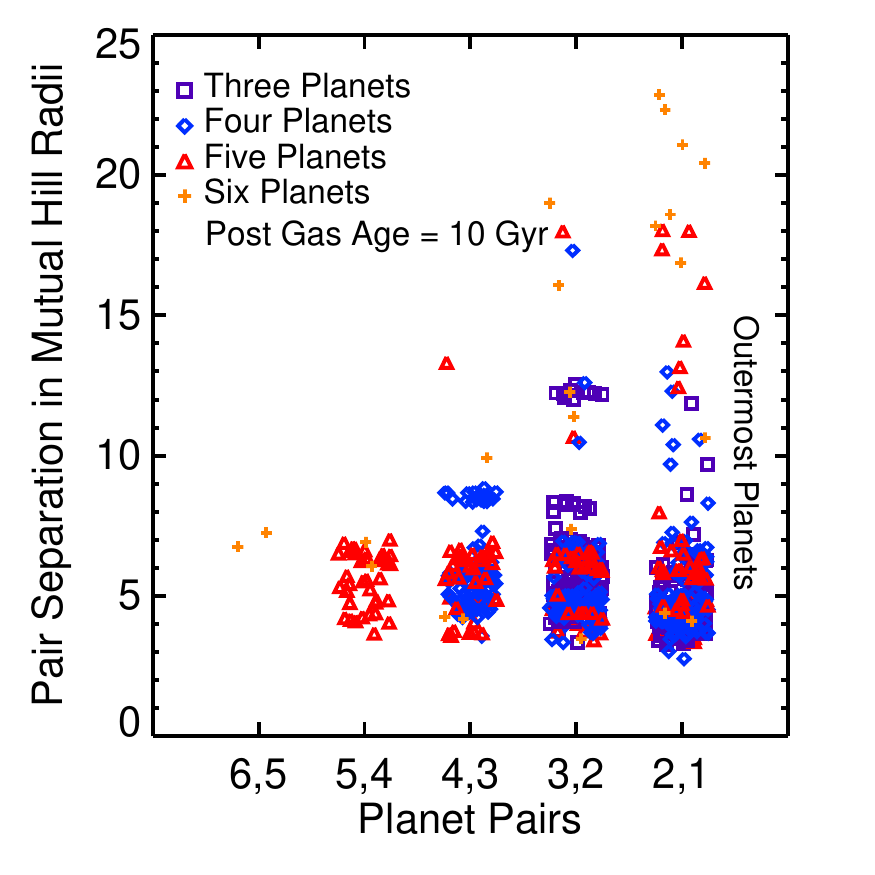}
\caption{\label{fig:hill_start}
The mutual Hill radii separation of adjacent planet pairs for the 3--30 AU simulations after the gas disk simulations are complete (left) and after 10 Gyr (right). X position within a column is slightly randomized for readability. Planet 1 is the farthest from the star. Planets are numbered upwards so a system with four planets has planets 1, 2, 3, and 4. After the 10 Gyr integration, many systems have far higher hill separations then they initially had, owing to the tendency of instabilities to greatly alter a system. If a planet is ejected, the number corresponding to a planet is updated an new comparisons are made. That is, if planet 1 is ejected, planet 2 is renamed to planet 1 and a pair is then determined between the new planet 1 and planet 0.}
\end{figure*}

We assess the impact of the initial post-gas spacing in mutual Hill radii and its relationship to stability of the 10 Gyr simulations. We present the starting (i.e., right after the gas disk phase) and final (i.e., after 10 Gyr) mutual Hill radii separations for adjacent planet pairs in Figure \ref{fig:hill_start}. At the start of the lifetimes all the planets have roughly equal separations of mutual Hill radii. Some sets display average mutual Hill radii values that differ from the majority of sets (see the blue diamond clump at 3,2 and the purple square clump at 2,1). These pairs tend to remain stable at these wider separations.

By comparing the two plots, we find that the planets furthest from the star (i.e., planets on the right) experience the greatest change in mutual Hill radii separation (especially amongst systems with four or more planets). Those pairs with Hill radii above 10 (right-side of Figure \ref{fig:hill_start}) are exclusively survivors of instability events (or pairs that began the post-gas stage with such wide separations, as mentioned in the previous paragraph). 

We find that if a system had an individual pair with a initial mutual Hill radii values smaller than 3.5, it had a greater likelihood of instability. However, these represented a small fraction of the total pairs. We conclude that initial mutual Hill radius-- within the narrow range encompassed in our starting conditions -- was not a primary driver of the stability rate for a system, although its final value can be used to assess which systems experienced instabilities.

\subsection{1 AU Planets}
\label{1au_plan}
The simulations by DD16 only included planets in a range from 3 to 30 AU (as based upon the protoplanetary disk observations, it is unclear if the deep and wide gaps extend to  within 3 AU). For simulation set, we run 10 new simulations with an additional planet near 1 AU (see Section \ref{section:methods} for details). The 1 AU planets still fulfill the transitional disk criteria presented by DD16: i.e., a planet massive enough to carve out a deep gap; packed closely enough with other planets to create a continuous gap; and close enough to the star to clear the disk at 1 AU. We run the simulations for 1 Myr the same gas damping parameters as the corresponding 3--30 AU systems.  Then we run the simulations for 1 Gyr post-gas. The simulation timescale is shorter than for the 3--30 AU systems to keep the run time feasible. We run 200 simulations in total, each with 4 to 7 planets.

The increased number of planets reduces the stability of the system. Several 1 -- 30 AU sets had stability rates similar to their 3--30 AU counterparts but a majority of the 1--30 AU sets saw substantial decreases in stability compared to their 3--30 AU counterparts. The reduction in stability is expected from past studies. For example,  \citet{multi_planet_stability} found that adding planets to a system led to shorter instability timescales. However, the impact was more modest for systems with higher multiplicity: for example, they found that a four planet system would go unstable significantly faster than a three planet system, but a seven planet system would only go unstable a little faster than a six planet system. We find that the 1--30 AU systems have shorter instability timescales, particularly for sets that originally contained 3 or 4 planets. The 5 and 6 planet sets were already frequently unstable by 1 Gyr: thus the additional instability from the extra planet has less impact on the final stability fraction.

\begin{figure}[ht]
\centering
\includegraphics[width=.45\textwidth]{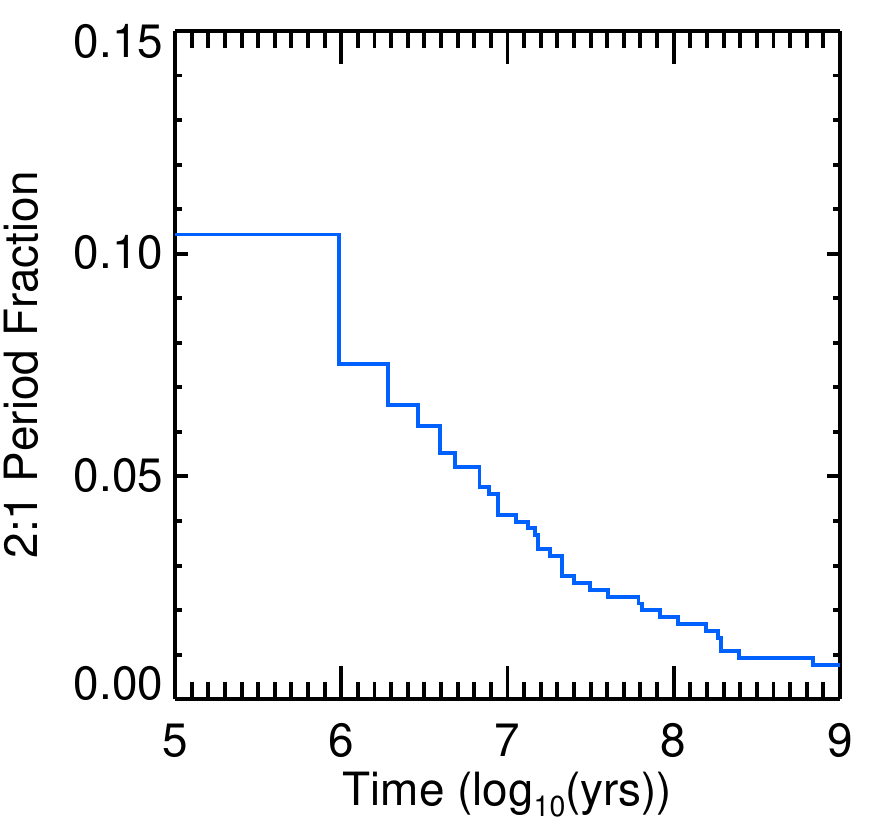}
\caption{\label{fig:stability_lifetime_histogram_1au_period}
A CDF showing the fraction of pairs with 2:1 MMR based on the period ratio for 1--30 AU systems. About 10\% of pairs begin in resonance
according to this metric, a slight decrease from the same analysis for 3--30 AU systems (Figure \ref{fig:stability_lifetime_histogram_period}). The two types of systems otherwise should identical trends.
}
\end{figure}

We present a CDF for pair-by-pair resonance status based on period ratio for the 1 -- 30 AU systems (Figure \ref{fig:stability_lifetime_histogram_1au_period}; in a fashion identical to the 3--30 AU systems in Figure \ref{fig:stability_lifetime_histogram_period}). The observed trends are mostly identical between the two types of systems, barring a slightly lower initial resonance fraction for the 1--30 AU systems (approximately 10\% compared to the prior 13\%). The fraction of 2:1 period pairs at 9 Gyrs is likewise approximately 2\% lower for the 1--30 AU systems. Adding a planet at 1 AU is unable to replicate the trend of younger ages for resonant pairs found by \citet{koriski}, which requires a typical disruption timescale $\sim 1--10$ Gyr and near 100\% initial fraction \citep{dong16}. Instead, our systems go unstable usually within $\sim 10--100$ Myr.

\section{Eccentricity}
\label{sec:eccentricity}
\begin{figure*}[ht]
\centering
\includegraphics[width=0.3\textwidth]{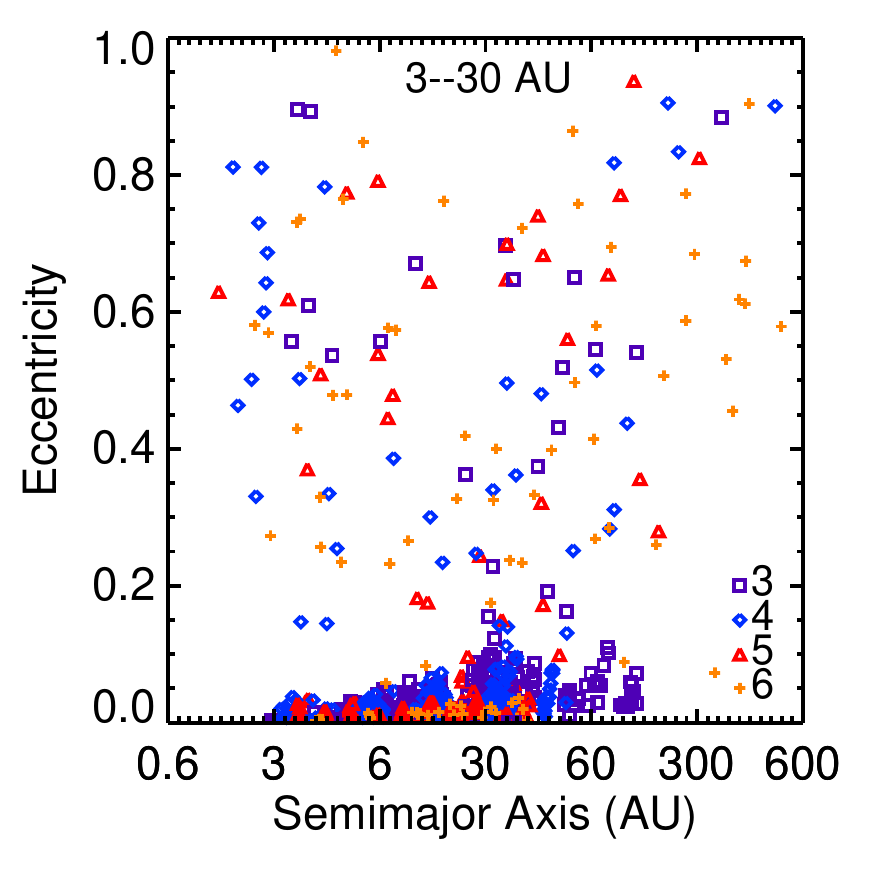}
\includegraphics[width=0.3\textwidth]{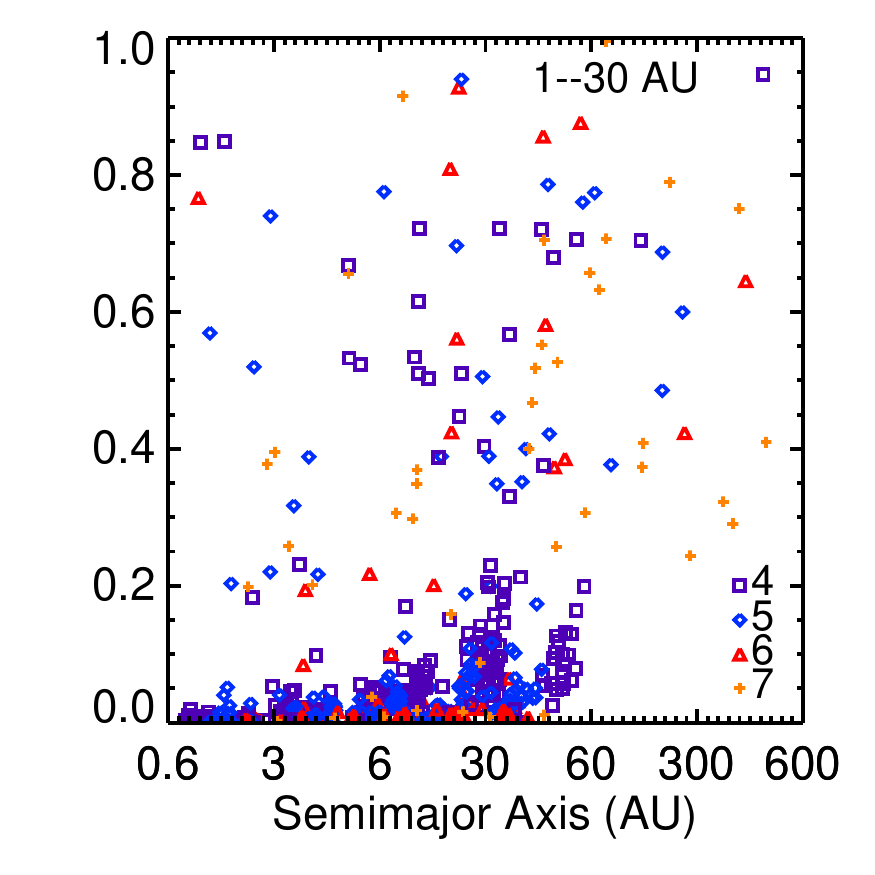}
\includegraphics[width=0.3\textwidth]{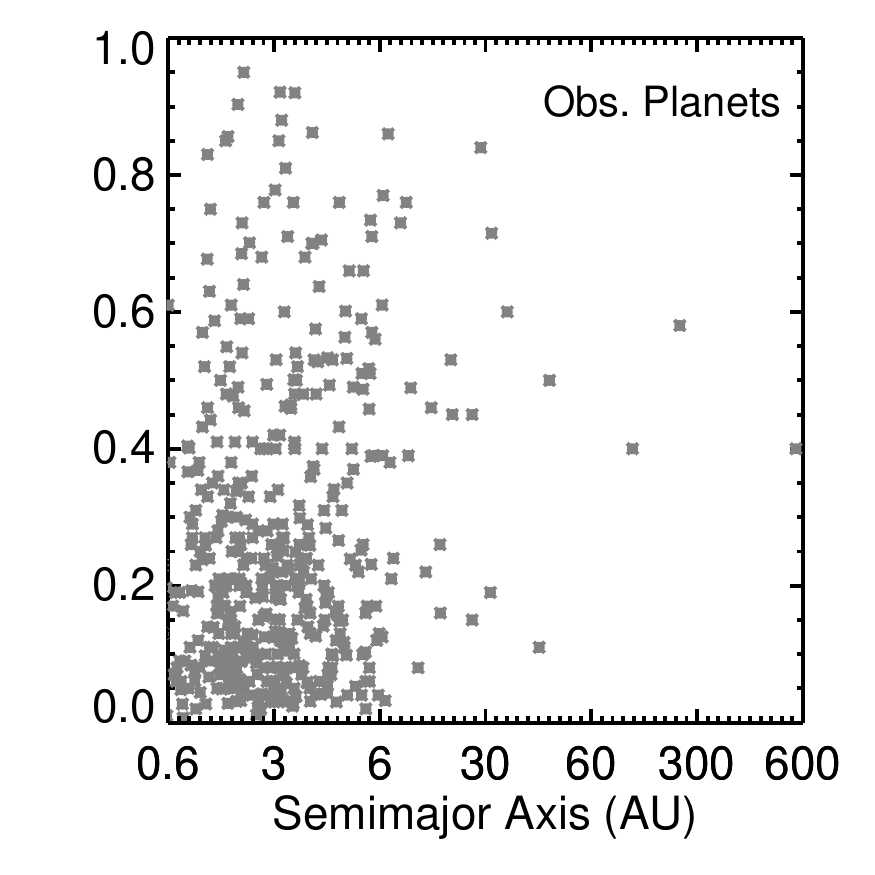}
\caption{\label{fig:a_vs_e_two_stable}
Eccentricity versus semimajor axis for the 3--30 AU systems at 1 Gyr (left), the 1-30 AU systems at 1 Gyr (middle), and a selection of observed exoplanets (right). Simulated systems are color coded by the initial number of planets. Most simulated systems remain at low eccentricity $e \lesssim 0.1$~for the entire evolution, although systems that began with more planets are more likely to undergo instabilities and end up on elliptical orbits.
}
\end{figure*}

\begin{figure}[ht]
\centering
\includegraphics[width=.4\textwidth]{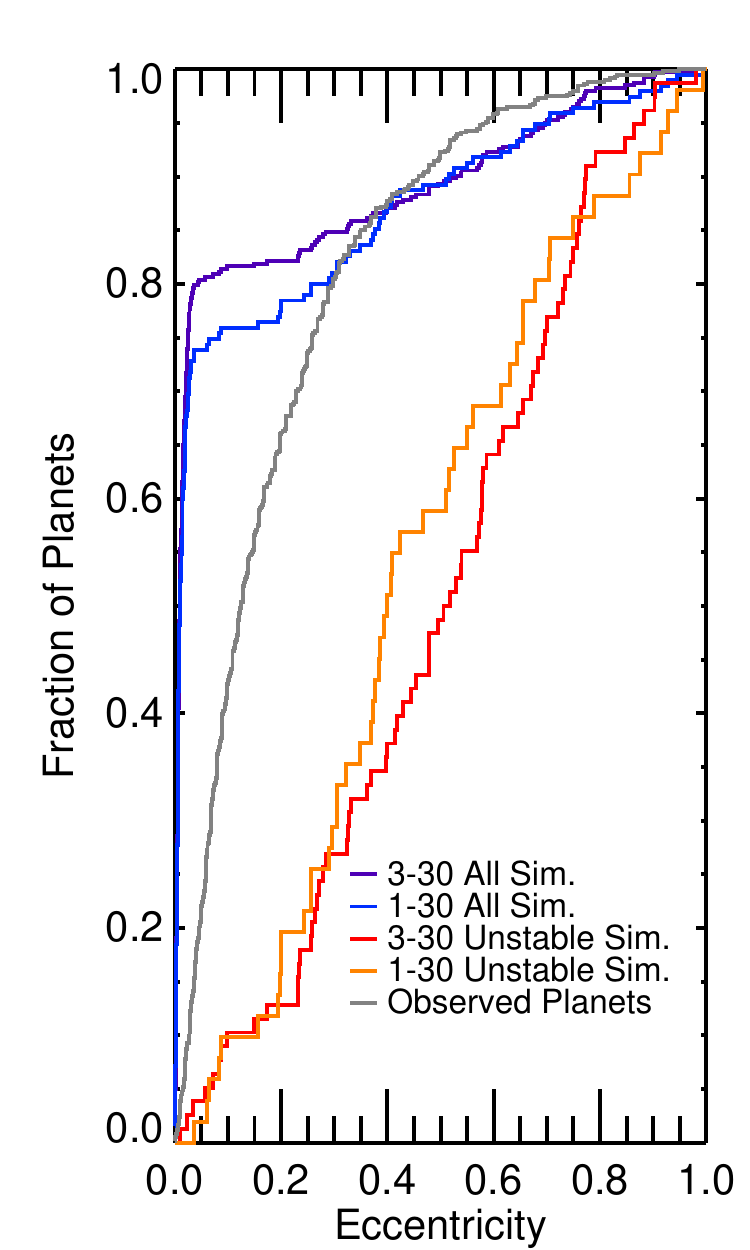}
\caption{\label{fig:eccen_all}
A CDF comparing the final eccentricities of five different system pools. All 3-30 AU results are computed at 1 Gyr to make them directly comparable to the 1-30 AU results. Unstable systems are systems where at least one planet experienced an instability event.
}
\end{figure}

The eccentricity distributions of our final systems is a relic of how the planets evolved with time. In Figure \ref{fig:a_vs_e_two_stable}, we plot the 1 Gyr eccentricity versus semimajor axis distribution for the 3--30 AU systems (left) and 1--30 AU systems (center). Furthermore, in Figure \ref{fig:eccen_all}, we plot a CDF for the 1 Gyr eccentricity of several categories. Of particular importance are the "unstable" categories for both the 3--30 AU and 1--30 AU systems: the unstable category shows the CDF of 1 Gyr eccentricities for surviving planets in systems that experienced instability events. Since extending the  3--30 AU systems to 10 Gyr only only decreased the fraction of circular orbits (<0.05 eccentricity) from 80\% to 70\%, we show the results at 1 Gyr to compare to the 1--30 AU systems, which were only simulated for 1 Gyr (Section \ref{1au_plan}).

We studied the influence of ejections and collisions on the eccentricity of surviving planets. In four and five planet systems (which had 80\%  and 50\% stability, respectively), the mean eccentricity of all survivors was 0.16 $\pm$ 0.25 while the eccentricity of survivors from unstable systems was 0.47 $\pm$ 0.23. As anticipated instability events increased the eccentricity of the survivors. Further trends were found in the type of instability events. Those systems which only experienced collisions had a final mean eccentricity of 0.33 $\pm$ 0.35 while those survivors that only experienced ejections had a final mean eccentricity of 0.50 $\pm$ 0.20. Although both types of instability event correlated with higher eccentricity, systems that experienced ejections saw a larger increase in the eccentricity of survivors.

The 3--30 AU systems and the 1--30 AU systems exhibit similar trends: nearly all planets with an eccentricity greater than 0.2 are in systems that experienced some instability event (primarily ejections, which occurred about 2.5 times more frequently than collisions) during the simulation. Both sets of simulations display a high concentration of low eccentricity planets: trending below 0.1 within 30 AU and below 0.2 beyond 30 AU. Including the additional planet near $\sim 1 au$ slightly increased the typical eccentricities of surviving planets, though the additional planet itself typically remains at very low eccentricity if it survives. 

We compare the eccentricities between our simulations and observed exoplanets in both Figure \ref{fig:a_vs_e_two_stable} (right) and Figure \ref{fig:eccen_all}. We use known Jovian mass planets (0.3 to 10 Jupiter M*$\sin(i)$) taken from the Exoplanet Archive on Aug. 4th, 2022. For our comparison sample, we select from the database those planets with eccentricities $> 0$. Planets discovered by various methods including radial velocity surveys and direct imaging are included in the comparison sample. Although there can be some biases in the measured values \citep{2010EAS....42..169P}, observers can often measure eccentricity for giant planets detected by radial velocity, which is how the vast majority of planets suitable for comparison (i.e., a similar mass and semimajor axis range) to our sample are detected.

The observed planets show eccentricity values with a concentration towards low eccentricity ($e < 0.15$). However, the simulations show a much more extreme concentration towards very low eccentricity ($e < 0.05$). Unstable configurations can reach high eccentricities (Figure \ref{fig:eccen_all}), but our  configurations (even if we limit to certain sets) apparently do not produce the correct mix of stable and unstable systems. Based on the large discrepancy between the simulations (even when considering only unstable simulations or only 5 and 6 planet systems) and the observed exoplanet eccentricities, we conclude that the systems created in DD16, while capable of creating transitional disks, can not reproduce evolved systems.

In contrast, other studies of planet-planet scattering that did not include a gas disk stage and/or require stability during the gas disk stage (e.g., \citealt{2008ApJ...686..580C,2008ApJ...686..603J}) produced eccentricity distributions that matched the observations well. Our use of equal-mass planets likely does not account for the difference, as such configurations are just as likely to lead to elliptical orbits, albeit with a narrow distribution \citep{2001Icar..150..303F}. Instead, our requirement that configurations remain stable during the gas disk stage to maintain a cavity \citep{dong16} apparently ensures too much stability to excite eccentricities. Our configurations do not achieve the ``dynamically active'' state identified by \citet{2008ApJ...686..603J} that erases the memory of initial conditions and are reminiscent of the ``dynamically cold'' configurations explored by \citet{daws16} that remain stable after the gas disk stage.

\section{Conclusions}
\label{sec:conclusion}
We simulated the long-term evolution of planetary systems capable of carving out and maintaining a transitional disk during the gas stage, using initial conditions from DD16. We subjected the planets to eccentricity damping during the disk stage and simulated the systems another 10 Gyr post-disk.

Our main finding (Section \ref{sec:eccentricity}) was that our systems tend to remain on stable, circular orbits. The typically very low eccentricities are at odds with those observed in real systems of giant exoplanets discovered via the radial velocity method. For example, fraction of planets with eccentricities between 0 and 0.2 was approximately 25\% higher in our simulated systems compared to real systems. Among our simulated systems that experience an instability, the eccentricities can reach the observed high values, but too few of our systems go unstable. The stability of systems showed a dependence on multiplicity:  three and four planet systems were more stable (i.e., experienced no collisions or ejections) compared to five and six planet systems by a significant margin (86\% stable versus 33\% stable, respectively). This trend is consistent with prior studies showing that higher planet multiplicity decreases stability  \citep{multi_planet_stability}. However, even when considering only high multiplicity systems and adding planets interior to those needed to carve observed gaps, our systems were too stable and circular.

We also found that the presence of a 2:1 MMR in a system -- which were commonly established in our simulations during the gas disk stage -- did not significantly impact the overall likelihood of going unstable over 10 Gyr. \citet{koriski} found that the presence of a 2:1 period ratio for a planetary pair indicates a system is younger on average. However, in contrast to \citet{koriski} who found that the typical lifetime of a 2:1 MMR is near 4 Gyr, we found that half of our pairs broke their 2:1 period ratio by 10 Myr. We noted a slightly higher instability rate for the near 2:1 MMR systems (those systems where the resonance angle still circled through all degrees but with a preference for a certain value) when considering resonance on a pair-by-pair basis (57\% stable compared to approximately 70\% stable for the resonance or no resonance systems). 
 
In future work, we could explore configurations of unequal mass planets, though past work has shown this is unlikely to significantly boost the resulting eccentricities \citep{2001Icar..150..303F}. It may be possible to further fine-tune our gas disk stage initial conditions (i.e., gas disk properties and planet spacing) to produce a higher fraction of post-gas unstable systems. However, more likely stability during the transitional disk stage precludes wide-spread instabilities after the gas disk disappears. If so, transitional disks may be typically caused by other processes besides giant planets, such as photoevaporation (e.g., \citealt{pigl21}) or compact configurations of super-Earths and mini-Neptunes in low viscosity disk \citep{fung2017}. Another possibility is that giant planets undergo convergent widescale migration at the very end of the transitional disk stage that packs them much closer together. As proposed by \cite{marel21}, this explanation would also help account for the large size of transitional disks cavities, compared to the peak in giant planet occurrence at smaller semi-major axes. However, it would need to be explored what could trigger this just in-time migration.

\acknowledgments
Computations for this research were performed on the Pennsylvania State University's Institute for Computational \& Data Sciences Advanced CyberInfrastructure (ICS-ACI). This content is solely the responsibility of the authors and does not necessarily represent the views of the Institute for CyberScience.
The Center for Exoplanets and Habitable Worlds is supported by the Pennsylvania State University and the Eberly College of Science.
This project was supported in part by NASA XRP NNX16AB50G and NASA XRP 80NSSC18K0355, the National Science Foundation under Grant No. NSF PHY-1748958, and the Alfred P. Sloan Foundation's Sloan Research Fellowship.
This research has made use of the NASA Exoplanet Archive, which is operated by the California Institute of Technology, under contract with the National Aeronautics and Space Administration under the Exoplanet Exploration Program.

\facility {Exoplanet Archive}
\software{Numpy}

\bibliographystyle{yahapj}
\bibliography{references}

\appendix
\section{Gas Stage Simulations}

Some of the gas stage simulations are taking directly from DD16 but, as described in Section 2, we run supplemental gas stage simulations as part of this course. Gas stage simulations use the {\tt mercury6} Bulirsch-Stoer hybrid integrator with a timestep of and Bulirsch-Stoer accuracy parameter of $10^{-12}$) \citep{1999MNRAS.304..793C}. Each configuration features three to six equal mass planets, with initial semimajor axes between 2 and 20 AU. The initial semi-major axes were selected so that the planets just barely open a common gap in the disk spanning 3--30 AU. Each configuration (Table \ref{tab:config}) meets this condition for an assumed disk viscosity, disk scale height, and planet mass (0.5--10 Jupiter masses). 

For each configuration, we run 20 random realizations; 10 were run as part of DD16 and 10 are new to this paper. The semi-major was randomized from the default values listed in Table \ref{tab:default_au} by about 5\%, drawing from a normal distribution centered on the default value. Initial eccentricities were 0. Initial inclinations were random set to $\sim 0.01^\circ$ to avoid perfectly coplanar planets. The initial mean longitude, longitude of periapse, and longitude of ascending node of each planet are randomized between 0 and $2\pi$.

The effects of the gas within the gap were implemented using prescriptions from \citet{papa00}, \citet{komi02}, \citet{ford07}, and \citet{rein12} as user-defined forces in {\tt mercury6}, following \cite{daws15} and DD16. The timescales for gas damping are:
\begin{eqnarray}
\label{eqn:damp}
\tau = 0.029 \frac{{\rm g\ cm}^{-2}}{\Sigma_{30}} \left(\frac{a}{\rm AU}\right)^2 \frac{M_\odot}{M_{\rm p}} {\rm yr} \times & \nonumber\\
& 1&, v < c_s \nonumber \\
&\left(\frac{v}{c_s}\right)^3&, v > c_s, i < c_s/v_{\rm kep} \nonumber  \\
&\left(\frac{v}{c_s}\right)^4&, i > c_s/v_{\rm kep} \nonumber \\
\end{eqnarray}
\noindent where $v = \sqrt{e^2+i^2} v_{\rm kep}$ and $v_{\rm kep}$ is the Keplerian velocity, and the sound speed $c_s = 1.29 {\rm km/s} \left(\frac{a}{\rm AU}\right)^{-1/4}$. We impose $\dot{e}/e = -1/\tau$ and $\dot{i}/i = -2/\tau$ \citep{komi02}. The value of $\Sigma_{30}$ for each configuration is listed in Table \ref{tab:config}. Following DD16, we do not impose migration ($\dot{a}$) or precession ($\dot{\varpi}$, $\dot{\Omega}$) because DD16 found that these effects were negligible except for fine-tuned values. Furthermore, the direction and magnitude of migration is sensitive to uncertain disk conditions; the magnitude is typically small in a depleted cavity. See Section 4.2.3 of DD16 for more detail.

\begin{table}
\centering
\caption{3 -- 30 AU Systems Default Semimajor Axis}
\footnotesize
\begin{tabular}{l|l|l|l|l|l|l|l|l}
\\
\hline
Name & $P_1$ $a$ & $P_2$ $a$ & $P_3$ $a$ & $P_4$ $a$ & $P_5$ $a$ & $P_6$ $a$ \\
& (AU) & (AU) & (AU) & (AU) & (AU) & (AU) \\
\hline
{\tt 3-5}   &   2.6 & 7.1 &     19.4 &- &- &-\\
{\tt 3-10a}	&   2.2 &    6.2   &    18.1 &- &- &-\\
 {\tt 3-10b}&   2.6  &   6.9    &   18.5  &- &- &-\\
 {\tt 3-10c}&   3.3   &   8.1    &  19.6  &- &- &-\\
 {\tt 3-10d}&   3.3   &  8.0    &  19.0  &- &- &-\\
 {\tt 3-10e}&   4.3  &    9.4  &      20.5  &- &- &-\\
 {\tt 4-2}	&   2.8   &    5.7     &   11.5 &   23.1  &- &-\\
{\tt 4-5a}	&   2.3 & 4.8  &   10.1 & 21.0 &- &- \\
{\tt 4-5b}	&   2.4 &  4.9 &  10.8 &    20.7  &- &-  \\
{\tt 4-5c}	&   2.5  &  5.0 &     10.4 &     21.3  &- &-  \\
{\tt 4-5d}	&   3.4 &    6.3 &     11.8 &    22.1 &- &-\\
{\tt 4-10b}	&   2.2 &   4.6 &    9.8 &     20.8 &- &-\\
 {\tt 5-1}	&   2.5 &   4.3 &    7.6 &    13.2     &  23.1 &-  \\       
{\tt 5-2a}	&   2.9 &     4.8 &  8.2 &  13.8 &  23.3 &- \\  
{\tt 5-2b}	&   3.7 &     5.8 &  9.3 &    14.9    &  23.8 &- \\   
 {\tt 5-5b}	&   2.5 &     4.3 &   7.5 &   13.1 &  22.9 &-    \\  
 {\tt 6-0.5}&   3.5 &     5.1 &     7.6 &    11.3 &   16.7 &   24.8   \\          
 {\tt 6-1}	&   3.6 &    5.3 &   7.8  &  11.4    & 16.8   & 24.8    \\              
\hline
\end{tabular}
\label{tab:default_au}
\tablecomments{The default semimajor axis ($a$) for each planet in a set, rounded to 0.1 AU.}
\end{table}

\end{document}